\documentclass[a4paper,twocolumn,amsmath,amssymb,aps]{revtex4-2}
\usepackage[utf8]{inputenc}
\usepackage[T1]{fontenc}
\usepackage[english]{babel}
\usepackage{hyperref}
\usepackage{graphicx}
\usepackage{dcolumn}
\usepackage{sidecap}
\usepackage{bm}
\usepackage{braket}
\usepackage{sistyle}

\DeclareMathOperator{\e}{e}
\newcommand{\I}{\mathrm{i}}

\usepackage{xcolor, soul} 
\usepackage{color}

\begin{document}

\title{Tight qubit uncertainty relations studied through weak values in \\neutron interferometry}
\author{Andreas Dvorak$^1$}
\author{Ismaele V. Masiello$^1$}
\author{Yuji Hasegawa$^{1,2}$}
\author{Hartmut Lemmel$^{1,4}$}
\author{Holger F. Hofmann$^3$}
\author{Stephan Sponar$^1$}
\affiliation{%
$^1$Atominstitut, TU Wien, Stadionallee 2, 1020 Vienna, Austria\\
$^2$Department of Applied Physics, Hokkaido University, Kita-ku, Sapporo 060-8628, Japan\\
$^3$Graduate School of Advanced Science and Engineering, Hiroshima University, Kagamiyama 1-3-1, Higashi Hiroshima, 739-8530, Japan\\
$^4$Institut Laue Langevin, 38000 Grenoble, France}
\date{\today}

\begin{abstract}
In its original formulation, Heisenberg’s uncertainty principle describes a trade-off relation between the error of a quantum measurement and the thereby induced disturbance on the measured object. However, this relation is not valid in general. An alternative universally valid relation was derived by Ozawa in 2003, defining error and disturbance in a general concept, experimentally accessible via a tomographic method. Later, it was shown by Hall that these errors correspond to the statistical deviation between a physical property and its estimate. Recently, it was discovered that these errors can be observed experimentally when weak values are determined through a procedure named ``feedback compensation''. Here, we apply this procedure for the complete experimental characterization of the error-disturbance relation between a which-way observable in an interferometer and another observable associated with the output of the interferometer, confirming the theoretically predicted relation. As expected for pure states, the uncertainty is tightly fulfilled.
\end{abstract}

\maketitle

\section{Introduction}
Heisenberg's uncertainty principle \cite{Heisenberg27} is without any doubt at the very heart of quantum physics. Nevertheless, several formulations coexist which address different physical scenarios or measures. Heisenberg's uncertainty principle formulated in terms of standard deviations is uncontroversial and demonstrated in various quantum systems. Its best known formulation is probably the product of the position and momentum standard deviations given by $\Delta(Q)\,\Delta(P)\geq\frac{\hbar}{2}$, which was rigorously proven by Kennard \cite{Kennard27} in 1927 and in 1929 generalized by Robertson \cite{Robertson29} to arbitrary pairs of non-commuting observables $\hat A$ and $\hat B$ expressed as $\Delta(\hat A)\,\Delta(\hat B)\geq\frac{1}{2}\braket{[\hat A,\hat B]}$. 

However, uncertainty relations in terms of standard deviations describe the limitation of preparing quantum objects and have no immediately obvious relevance to the limitation of measurements on single systems, as originally suggested by Heisenberg in the beginning of his paper \cite{Heisenberg27}. Heisenberg's starting point is a relation between the precision of a position measurement and the disturbance it induces on a subsequent momentum measurement of a particle -- more precisely of an electron. This is beautifully captured in the famous $\gamma$-ray microscope thought experiment, which is solely based on the Compton-effect: \emph{At the instant when the position is determined -- therefore, at the moment when the photon is scattered by the electron -- the electron undergoes a discontinuous change in momentum. This change is the greater the smaller the wavelength of the light employed that is, the more exact the determination of the position} \cite{Heisenberg27}.

A relation is given as $q_1\, p_1 \sim h$, for the product of the \emph{mean error} $q_1$ of a position measurement (error) and the discontinuous change (disturbance) $p_1$ of the particle's momentum. Heisenberg's original formulation can be read in terms of a modern treatment of quantum mechanics as $\varepsilon(Q)\,\eta(P)\geq\frac{\hbar}{2}$, where $\varepsilon(Q)$ is the error of a measurement of the position observable $Q$ and $\eta(P)$ is the disturbance of the momentum observable $P$.

The generalized form of Heisenberg's original error-disturbance relation, for arbitrary pairs of observables $\hat A$ and $\hat B$, would read $\varepsilon(\hat A)\,\eta(\hat B)\geq\frac{1}{2}\braket{[\hat A,\hat B]}$. However, such a naive generalization of a Heisenberg-type error-disturbance relation for arbitrary observables is not valid in general \cite{Arthurs88,Ozawa91}. 
In 2003 Ozawa introduced a more accurate formulation of the error-disturbance relation for generalized measurements, based on a rigorous theoretical analysis of the measurement process \cite{Ozawa03}. In addition to the error $\varepsilon$ and the disturbance $\eta$, this bound also includes the standard deviations $\Delta$ of the initially prepared state.
Details are presented in Sec.\,II. A different approach in terms of a trade-off relation for errors and disturbance in quantum measurements was presented recently by Busch and his co-workers \cite{Busch13}, aiming to maintain the original form of Heisenberg-type error-disturbance uncertainty relation by appropriate definitions of error and disturbance in terms of differences between output distributions.  However, there continues to be debates as to the appropriate measure of measurement (in)accuracy and of disturbance \cite{Ozawa03,OzawaPLA03,Hall04,Branciard13,Busch13,Busch13PRA,Busch14,Buscemi14,Barchielli18,Pan19}.

Initially it was assumed, that the Ozawa-Hall errors $\varepsilon$ \cite{Ozawa03,Hall04} have no experimentally observable properties \cite{Werner04}. However, in recent years experiments reconstructed the uncertainties based on statistical assumptions that were motivated by a theoretical analysis of the formalism, either using a tomographic reconstruction, i.e., three-state-method \cite{Ozawa04,Erhart12,Edamatsu13,Ringbauer14,Sponar17}, or weak measurements \cite{Lund10,Steinberg12}. Furthermore, it has been recently predicted in \cite{Hofmann21} that Ozawa-Hall uncertainties can be directly observed as the uncertainty in the rotation of a probe qubit when the method of feedback compensation is used. An experimental demonstration of feedback compensation applied to neutron interferometry is reported in \cite{Lemmel2022}, where cases of purely real weak values have been studied. 

In this paper, we apply the feedback compensation method to extract the real part of complex weak values for determining the error. Furthermore, all other parts of the Ozawa uncertainty relation are measured and compared with the theory. We show that the uncertainty relation is tightly fulfilled for various initial interferometer states. The remaining paper is organized as follows: In Sec. II, we introduce the theory of the Ozawa uncertainty. In Sec. III, we describe the setup and present the measurement results. In Sec. IV we discuss the results.

\section{Theory}
\subsection{General framework}
In Ozawa's \emph{operator-based} approach \cite{Ozawa03}, or \emph{operator formalism}, the measurement  process is described by an indirect measurement model, introduced in \cite{Ozawa88}.
Let $\bf{S}$ be a system described by the Hilbert space $\mathcal{H}$ (\emph{object system}) and initial state $\ket{\psi}$. The operator $\hat A$ represents the observable to be measured and the (in general non-commuting) operator $\hat B$ the observable which is potentially disturbed by the measurement. Let $\bf{P}$ be a \emph{probe system} described by the Hilbert space $\mathcal{K}$, state $\ket{\xi}$ and meter observable $\hat M$. 
Then the {\em indirect measurement} apparatus $\bf{A}$ is defined by the quadruple $(\mathcal{K},\ket{\xi},\hat U,\hat M)$ with the unitary operator $\hat U$ on $\mathcal{H}\otimes\mathcal{K}$ describing the time
evolution of the composite system $\bf{S}+\bf{P}$ during the measuring interaction. Here we treat only the case where the meter observable $\hat M$ has non-degenerate eigenvalues. In this case, $\hat M$ has a spectral decomposition
$\hat M=\sum_{m}m\ket{m}\bra{m}$, where $m$ varies over eigenvalues of $\hat{M}$. Then, the apparatus $\bf{A}$ has a family $\{\hat M_{m}\}$ of operators, called
the {\em measurement operators}, defined by $\hat M_m=\bra{m}\hat U\ket{\xi}$. The {\em root-mean-square (rms) error} $\varepsilon(\hat A)$ of $\bf{A}$
for measuring an observable $\hat A$ of $\bf{S}$ and the {\em rms disturbance} $\eta(\hat B)$ imposed by $\bf{A}$ on an observable $\hat B$ of $\bf{S}$ are defined as
\begin{eqnarray}
    \varepsilon(\hat A)&=&\|\,(\hat U^{\dagger}({1\!\!1}\otimes \hat M)\hat U-\hat A\otimes \hat{1\!\!1})\ket{\psi}\ket{\xi}\|, \nonumber\\
    \eta(\hat B)&=&\|\,(\hat U^{\dagger}(\hat B\otimes {1\!\!1})\hat U-\hat B\otimes \hat{1\!\!1})\ket{\psi}\ket{\xi}\|,
    \label{eq:ErrDist_def}
\end{eqnarray}
 with $\vert\vert X\vert\psi\rangle \vert\vert=\langle\psi\vert X^\dagger X\vert\psi\rangle^{1/2}$.
Hence, the operator formalism defines a relation between the measurement outcome and the target observable. Then, it is proved \cite{Ozawa03,Ozawa04} that Ozawa's general uncertainty relation, given by
\begin{equation}
    \varepsilon(\hat A)\,\eta(\hat B)+\varepsilon(\hat A)\,\Delta(\hat B)+\Delta(\hat A)\,\eta(\hat B) \geq \frac{1}{2}\left\vert\bra{\psi}\left[\hat A,\hat B\right]\ket{\psi}\right\vert,
    \label{eq:Ozawa}
\end{equation}
holds for any state $\ket{\psi}$ of $\bf{S}$ and any indirect measurement model $(\mathcal{K},\ket{\xi},\hat U,\hat M)$.
Using $\{\hat M_{m}\}$ we can rewrite error and disturbance starting from their definitions in Eq.\,(\ref{eq:ErrDist_def}) as $\varepsilon^2(\hat A)=\sum_{m} \|\hat M_{m} ({m}-\hat A)\ket{\psi}\|^2$ and $\eta^2(\hat B)=\sum_{m} \|[\hat M_{m}, \hat B]\ket{\psi}\|^2$. If the ${\hat M_{m}}$ are mutually orthogonal projection operators, and using the Pythagorean theorem the error yields
\begin{equation}
    \varepsilon^2(\hat A) = \bra{\psi}\left(\hat A-\sum_m m\,\hat M_m\right)^2\ket{\psi}.
\end{equation}

It was shown by Hall \cite{Hall04} that this error corresponds to the uncertainty of a set of estimate values $A_m$ for a measurement performed in base $\ket{m}$
\begin{align}
    \varepsilon^2(\hat A) &= \bra{\psi}\left(\hat A-\sum_{m}A_{m} \ket{m}\bra{m}\right)^2\ket{\psi} \label{eq:error_proj}\\
    &=\langle \hat A^2\rangle+\sum_m p_m\Big(\big(A_m-\Re(\omega_m)\big)^2-\Re(\omega_m)^2\Big)\label{eq:ErrHallOrg}   
\end{align}
where $\langle \hat A^2\rangle=\langle \psi |\hat A^2|\psi\rangle$ denotes the expectation value of the squared operator, $p_m = |\langle \psi|m\rangle|^2$ denotes the probability for measurement outcome $m$ and $\omega_m$ denotes the weak value defined as
\begin{equation}
    \omega_m = \frac{\braket{m\vert \hat A\vert\psi}}{\braket{m\vert\psi}}.
\end{equation}
Hall further pointed out that the optimal estimates which minimize the uncertainty, cf. Eq. (\ref{eq:ErrHallOrg}), are given by the real part of the weak value.
\begin{equation}
    A^{\rm{opt}}_{m}=\Re(\omega_m)
\end{equation}
Then Eq. (\ref{eq:ErrHallOrg}) simplifies to
\begin{align}\label{eq:ErrExpVal}
  \varepsilon^2(\hat A)=\langle \hat A^2\rangle-\sum_m p_m(A^{\rm opt}_m)^2.
\end{align}
Using $\langle\hat A\rangle = \langle\psi|\hat A|\psi\rangle = \sum_m p_m \Re(\omega_m)$ it can be shown that Eq. (\ref{eq:ErrExpVal})
is equivalent to
\begin{align}\label{eq:ErrStdDev}
  \varepsilon^2(\hat A) &= \Delta^2(\hat A)-\sum_m p_m\left(A^{\rm{opt}}_m-\langle\hat A\rangle\right)^2
\end{align}
which relates the uncertainty to the squared standard deviation $ \Delta^2(\hat A) = \langle\hat A^2\rangle - \langle\hat A\rangle^2$ of the initial state.
Both Eq. (\ref{eq:ErrExpVal}) and Eq. (\ref{eq:ErrStdDev}) show that the Ozawa-Hall uncertainties can be determined from initial state expectation values $\langle \hat A\rangle$ and the optimal values of the estimates $A_m^{\rm opt}$. Although instructive, these relations might produce negative results in practise if the values of the estimates carry experimental errors. However, a more robust expression can be derived directly from Eq. (\ref{eq:error_proj}).
If the base  $\ket{m}$ is complete and orthogonal and the operator $\hat A$ self-adjoint Eq. (\ref{eq:error_proj}) turns into
\begin{alignat}{1}
    \varepsilon^2(\hat A)
    &= \sum_m \langle\psi| \left(\hat A^\dagger - A_m \right) |m\rangle\langle m| \left(\hat A-A_m\right)|\psi\rangle \nonumber\\
    & =\sum_m p_m\, |\omega_m -A_m|^2.\label{eq:ErrHall}
\end{alignat}
In this form the uncertainty is given by the weighted sum over the differences between estimates and weak values. Note that the weak values $\omega_m$ are in general complex values while the estimates $A_m$ are real ones. Consequently, the uncertainty will vanish if and only if the estimates are optimal and the weak values are real.

\subsection{States and observables}
Let us consider a neutron traversing a Mach-Zehnder interferometer. After the first \emph{tunable} beam splitter, the particle is put in a superposition of path states $\ket{1}$ and $\ket{2}$ of the form:
\begin{align}
    \ket{\psi(\chi)} = a_1\ket{1}+a_2\e^{\I\chi}\ket{2},\label{eq:initialstate}
\end{align}
where $a_1$ and $a_2$ are arbitrary (real) path amplitudes following ${a_1}^2+{a_2}^2=1$, and $\chi$ is a relative phase between the path states. We refer to $\ket{\psi(\chi)}$ as our initial state. Before exiting the interferometer, the path states are recombined using a 50:50 beam splitter which projects onto the states:
\begin{align}
    \ket{+} &= \frac{\ket{1}+\ket{2}}{\sqrt{2}}, \nonumber\\
    \ket{-} &= \frac{\ket{1}-\ket{2}}{\sqrt{2}}.
    \label{eq:eigenstates}
\end{align}
We refer to $\ket{+}$ and $\ket{-}$ as our final states. The probabilities of finding a neutron in the output ports are given by
\begin{align}
    p_\pm = \left\vert\braket{\pm\vert\psi(\chi)}\right\vert^2  = \frac{1}{2}\pm a_1\,a_2\cos\chi\,.\label{eq:probability}
\end{align}
In this experiment, the observable $\hat{A}$ corresponds to
\begin{align}
\hat{A}=\hat{\Pi}_1=\ket{1}\bra{1},
\end{align}
with $\hat{\Pi}_1$ being a projector representing the presence of a neutron in path 1. The observable $\hat{B}$ corresponds to
\begin{align}
    \hat{B} = \hat\sigma_x = \ket{+}\bra{+}-\ket{-}\bra{-}\label{eq:sigmax}
\end{align}
with $\hat\sigma_x$ being the operator relative to the output ports of the interferometer.

 Hence, the path eigenstates $\vert 1\rangle$ and $\vert 2\rangle$ span the Hilbert space of the object system $\bf{S}$. Furthermore the measurement operators are represented by the projectors onto the final states $\hat M_m=\vert\pm\rangle\langle\pm\vert$. For the probe system $\bf{P}$ we use the neutron's spin degree of freedom. 

\subsection{Ozawa uncertainty limit for path information in the interference measurement}
The uncertainty relations introduced by Ozawa provide an input state dependent lower bound for the error $\varepsilon(\hat A)$, disturbance $\eta(\hat B)$, standard deviations $\Delta(\hat A)$ and $\Delta(\hat B)$ of a successive measurement of observables $\hat{A}$ and  $\hat{B}$, expressed by Eq.\,(\ref{eq:Ozawa}).Since in our special case the observables $\hat M_{m}$ and $\hat B$ commute ($\left[\ket{\pm}\bra{\pm},\hat\sigma_x\right]=0$) the disturbance, given by $\eta^2(\hat B)=\sum_{m} \|[\hat M_{m}, B]\ket{\psi(\chi)}\|^2$, vanishes and the relation is reduced to
\begin{align}\label{eq:tightUVUR}
    \varepsilon(\hat A)\,\Delta(\hat B) \geq \frac{1}{2}\vert\bra{\psi(\chi)}[\hat A,\hat B]\ket{\psi(\chi)}\vert.
\end{align}
We experimentally measure both sides of Eq.\,(\ref{eq:tightUVUR}). For pure states, theory predicts that the Ozawa uncertainties are a {\em tight} bound. In practise, additional noise increases the value of $\varepsilon(\hat A)$ (LHS) and decrease the uncertainty bound (RHS) by decreasing the sensitivity of $\braket{\hat B}$ to small phase shifts.

\subsubsection{Error $\varepsilon(A)$,\, path presence $\omega_{1\pm},\,$ and feedback compensation in the interference measurement}

Applying Eq.\,(\ref{eq:ErrHall}) to our experiment we obtain the error
\begin{alignat}{1}
    \varepsilon^2(\hat{\Pi}_1)
    & =\sum_\pm p_{\pm}\left\vert \omega_{1\pm} - A_\pm\right\vert^2,
    \label{eq:error_proj2}
\end{alignat}
where the weak values $\omega_{1\pm}$ are given by
\begin{align}
  \omega_{1\pm} &= \frac{\braket{\pm\vert\hat\Pi_1\vert\psi(\chi)}}{\braket{\pm\vert\psi(\chi)}} = \frac{1}{1\pm\frac{a_2}{a_1}\e^{\I\chi}}.
  \label{eq:WeakValue}
\end{align}

\begin{figure}[!t]
    \centering
    \includegraphics[width=0.48\textwidth]{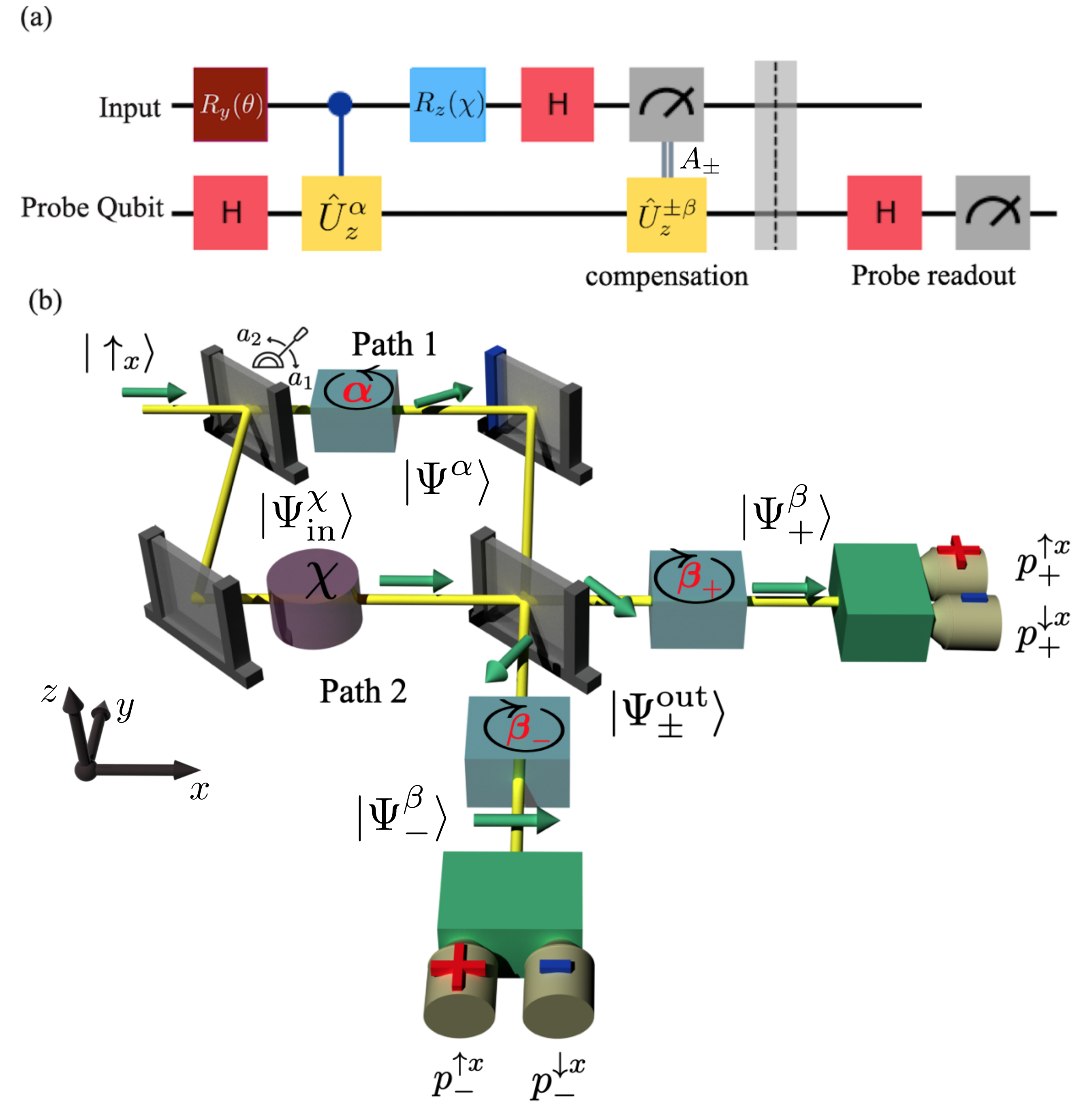}
    \caption{Scheme of \emph{feedback compensation} (a) from \cite{Hofmann21} as applied to a Mach-Zehnder interferometer (b).  After a coupling $\hat U^\alpha_{z}$ between object (interferometer paths) and probe system (spin), a compensation $\hat U_z^{\beta\pm}$ dependent on the output channel is applied, searching for the maximal value of $p_\pm^{\uparrow x}$ of the probe qubit in output $\ket{+}$ and $\ket{-}$, respectively. See Appendix D and E for details of the state evolution and calculation of observed intensities, respectively.}
    \label{fig:scheme}
\end{figure}

The {\emph{experimentally measurable}} output probabilities are given by $p_\pm=\vert\braket{\pm\vert\psi(\chi)}\vert^2=\frac{1}{2}\pm a_1\,a_2\cos\chi$. The optimal estimates $A^{\text{opt}}_\pm$ after the interference measurements are given by the real part of the weak value of the operator $\hat A=\hat\Pi_1$:
\begin{align}
  A^{\text{opt}}_\pm &=\Re (\omega_{1\pm}) = a_1^2\mp\frac{a_1 a_2\cos{\chi}}{1\pm2\,a_1\,a_2\cos{\chi}}\left(a_1^2-a_2^2\right).\nonumber\\
  \label{eq:opt}
\end{align}
These optimal estimates are determined experimentally using the feedback compensation scheme \cite{Lemmel2022}. 
The experimental setup is shown in Fig.\,\ref{fig:scheme}. The observable of interest -- here the presence of the neutron in path 1, mathematically represented by the projection operator $\hat\Pi_1$ -- weakly couples to a probe qubit through a controlled phase gate $\hat U_z^\alpha$. Experimentally, this is achieved by a small spin rotation in path 1 only. As a result of this interaction, the  spin rotation angle encoded in the probe qubit is proportional to the physical value of the projector onto path 1, where statistical fluctuations in this value result in a corresponding uncertainty of the rotation angle. If an estimate $A_\pm$ of the correct value (the ``path presence'') is available, a unitary operation $\hat U_z^{\beta\pm}$ in the exit paths can compensate the interaction and reduce the uncertainty of the rotation angle. Here, the estimated value corresponds to the ratio of the compensation rotation $\beta_\pm$ and the rotation of $\alpha$, i.e.
\begin{align}
A_\pm=\frac{\beta_\pm}{\alpha}.
\end{align}
An optimal compensation not only minimizes the error but also restores the original state of the probe qubit to a best possible level. By varying $\beta_\pm$ and observing the spin states in the interferometer outputs it is therefore possible to determine the optimal estimates  $A_\pm$ experimentally \cite{Lemmel2022} (see Appendix D for details on the optimal compensation angle). 

\subsubsection{Standard deviation $\Delta(B)$}
In our experiment $\hat B=\hat \sigma_x$ of the path qubit and with the expressions given in Eqs.\,(\ref{eq:initialstate}) to (\ref{eq:sigmax}) it is straightforward to calculate
\begin{align}
    \Delta(\hat B) &=\Delta(\hat\sigma_x) = \langle \hat\sigma_x^2\rangle - \langle \hat\sigma_x\rangle^2 = \nonumber \\ &=\sqrt{1-\left(2\,a_1\,a_2\cos\chi\right)^2} = 2\sqrt{p_+\,p_-}\,\,\,.
\end{align}
Thus, the probabilities $p_+$ and $p_-$ as function of $\chi$ can be determined directly from the interference fringes measured at the $+$ and $-$ output port respectively. 

\subsubsection{Lower bound of uncertainty relation (RHS)}
The path observable $\hat{A}$ appears in these relations as the generator of the phase shift,
\begin{align}
    \ket{\psi(\chi)} = \e^{\I\chi}\e^{-\I\hat{A}\chi}\ket{\psi(0)}.
\end{align}
It is therefore possible to identify the commutation relation of $\hat{A}$ and $\hat{B}$ with the phase dependence of the expectation value of $\hat{B}$,
\begin{align}
    \frac{\text{d}}{\text{d}\chi}\braket{\hat{B}} = \I\braket{[\hat{A},\hat{B}]}.
\end{align}
It is possible to evaluate the commutation relation by using the gradient of the interference fringe. This may require measurements of very similar phases, especially where the gradient is very close to zero. For the Ozawa uncertainty limit, the expectation value of the commutation relation can then be replaced by the gradient of the interference fringe,
\begin{align}
    \frac{1}{2}\left\vert\braket{[\hat{A},\hat{B}]}\right\vert = \frac{1}{2}\left|\frac{\text{d}}{\text{d}\chi}\left(p_+-p_-\right)\right\vert.
    \label{eq:limit}
\end{align}
The measurement is always a projection on eigenstates of $\hat{B}$, so the disturbance $\eta_B$ is always zero. The commutation relation therefore describes a lower bound of the measurement error for any measurement results $A_+$ and $A_-$ assigned to the outcome of the $\hat{B}$ measurement. If $\hat{A}$ is the generator of dynamics, the commutation relation expresses the rate of change of the expectation value $\braket{\hat{B}}$ caused by the dynamics. The information about the value of $\hat{A}$ is therefore limited by the sensitivity of $\braket{\hat{B}}$ to dynamics generated by $\hat{A}$,
\begin{align}
    \varepsilon(\hat A)\,\Delta(\hat B) \ge \frac{1}{2}\left\vert\frac{\text{d}}{\text{d}\chi}\braket{\hat{B}}\right\vert.
\end{align}
This relation is a special case of the more general relation between phase sensitivity and Ozawa uncertainties of the generator \cite{Hofmann_2011}. The Ozawa uncertainty of the generator of a phase shift defines an upper bound of the Fisher information for that measurement and the actual phase sensitivity obtained in the measurement defines a lower bound of the Ozawa uncertainty.

Since it is experimentally very difficult to measure very similar phases, especially
where the gradient is very close to zero, it is better to determine $\frac{1}{2}\vert\braket{[\hat{A},\hat{B}]}\vert$ from $p_+$ by shifting the phase as $\chi\rightarrow\chi\mp\frac{\pi}{2}$. Namely (see Eq.\,\ref{eq:appRHS} in Appendix B for details)
\begin{align}
    \frac{1}{2}\left\vert\bra{\psi}[\hat{A},\hat{B}]\ket{\psi}\right\vert = \frac{1}{2}\left\vert\,p_+\left(\chi-\frac{\pi}{2}\right)-p_+\left(\chi+\frac{\pi}{2}\right)\right\vert.
    \label{eq:RHS}
\end{align}

\section{Experiment}
\subsection{Experimental Setup}
\begin{figure}[!h]
    \includegraphics[width=0.48\textwidth]{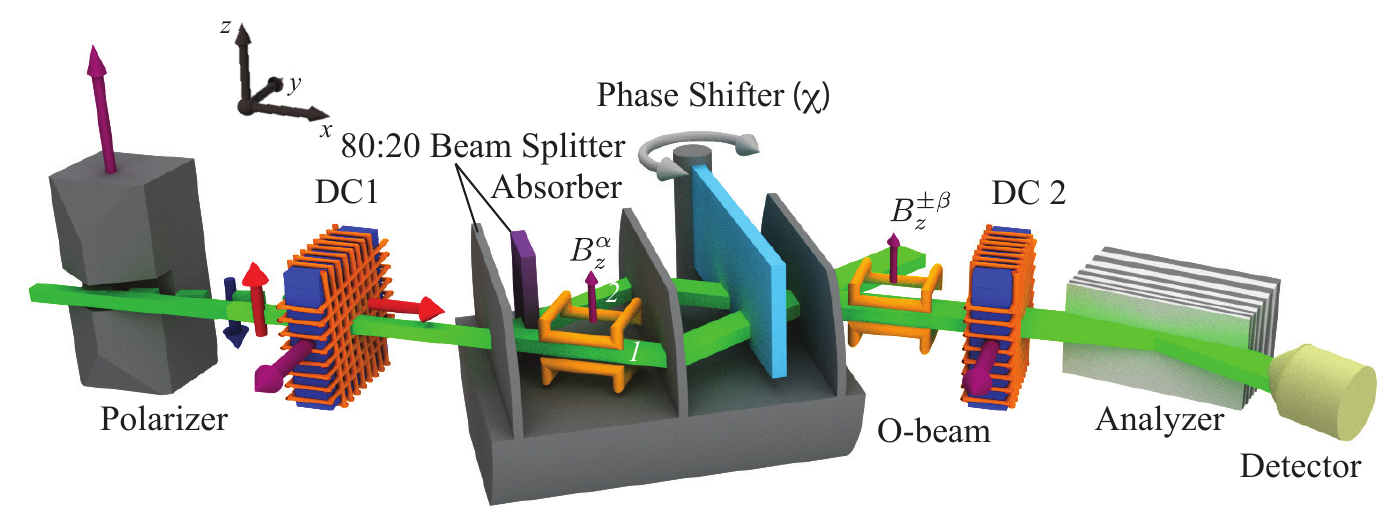}
    \caption{Experimental setup: polarized monochromatic neutrons enter the interferometer and are split into paths $\ket{1}$ and $\ket{2}$ at the first interferometer plate at a ratio of 1:4. Before the interferometer, the probe qubit is prepared  by a $\frac{\pi}{2}$ direct-current spin rotator  (DC\,1). In path $1$ the spin is rotated by an angle $\alpha$. The phase shifter adjusts the relative phase $\chi$ of the initial state $\ket{\psi(\chi)}$. Behind the interferometer (in the O-beam), the \emph{compensation} is applied, that is a spin counterrotation by angle $\beta_\pm$, dependent of the respective measurement context. The spin is analyzed in $\pm x$ direction by the combination of a $\frac{\pi}{2}$ direct-current spin rotator (DC\,2) and the magnetic supermirror. The neutrons are counted in a $^3$He detector.}
    \label{fig:setup}
\end{figure}

The experimental realization of the operations are depicted in Fig.\,\ref{fig:setup}. The neutrons are polarized by a magnetic prism which deflects the spin-down neutrons out of the Bragg acceptance angle of the interferometer crystal. The spin rotator DC1 rotates the remaining spin-up neutrons by $\frac{\pi}{2}$ into the initial $\ket{\uparrow_x}$ state. The spin rotator consists of a DC coil which creates a magnetic field pointing in the $-y$ direction. The spin precesses about the $y$ axis due to Larmor precession within the DC coil. In order to preserve polarization, a constant magnetic field (guide field) is applied throughout the setup in the $z$-direction causing the spin to precess around the $z$-axis. For the sake of graphical clarity, the guide field is not depicted in Fig.\,\ref{fig:setup}. The spin rotations $\alpha$ and $\beta$, accounting for operations $\hat U^\alpha_{z}$ and $\hat U_z^{\beta\pm}$, respectively, are realized by small Helmholz coils which modify the external overall $B_z$ guide field such that the spin precession in the $x$-$y$-plane is tuned. The precession angle is given by $-\frac{2\mu B^j_z\tau}{\hbar}$, where $\tau$ is the neutron’s transit time in the magnetic field region, with $j=\alpha,\beta$. The $\beta$ compensation and the spin analysis is realized only in the forward exit beam (O-beam), corresponding to the $\ket{+}$ state. The results for the $\ket{-}$ state can be retrieved at the forward beam by shifting the relative phase as $\chi\rightarrow\chi\pm\pi$.

The spin analysis is realized by a spin-dependent reflection from a Co-Ti supermirror array. The magnetic supermirror array consists of a stack of slightly bent glass plates coated with alternating layers of (magnetic) Cobalt and (non-magnetic) Titanium embedded in the vertical field of permanent magnets. The combination of materials is chosen such that the sum of the nuclear scattering length and the magnetic scattering length of Cobalt for one spin component equals the scattering length of Titanium. Then the layer structure is invisible for this spin component and it will not be reflected. Consequently, the supermirror only reflects the $\ket{\uparrow_z}$ state and the $\ket{\downarrow_z}$ state is absorbed. In combination with a $\frac{\pi}{2}$ spin rotator (DC2) it analyzes the $\ket{\uparrow_x}$ state required here. The position of the DC2 coil is adjusted in beam direction to catch the precessing spin at the correct angle required for the $\frac{\pi}{2}$ rotation.

\begin{figure}[!b]
    \centering
    \includegraphics[width=0.47\textwidth]{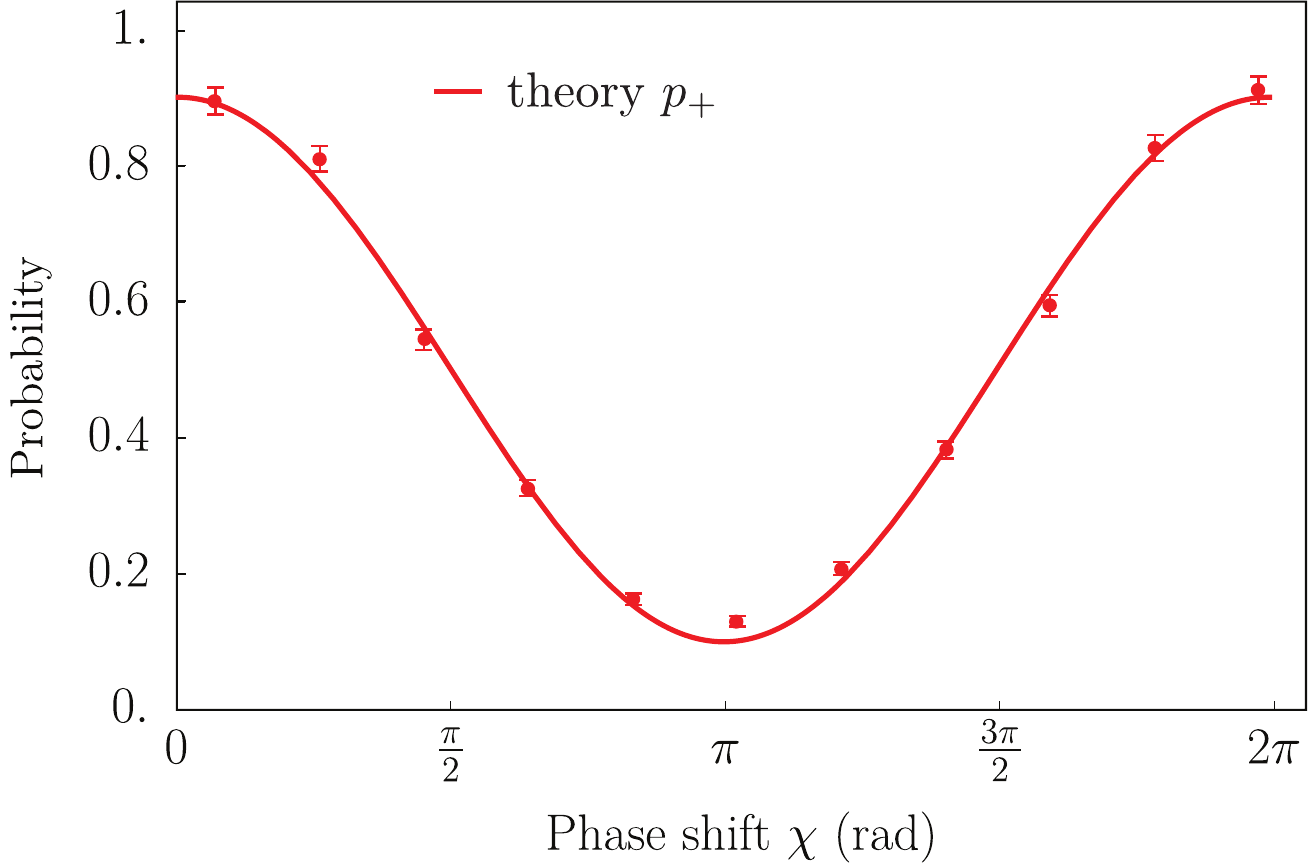}
    \caption{Measured probability $p_+$, vs. phase of initial state $\chi$, together with theoretical predictions.}
    \label{fig:p+}
\end{figure}

\begin{figure}[!b]
    \centering
    \includegraphics[width=0.44\textwidth]{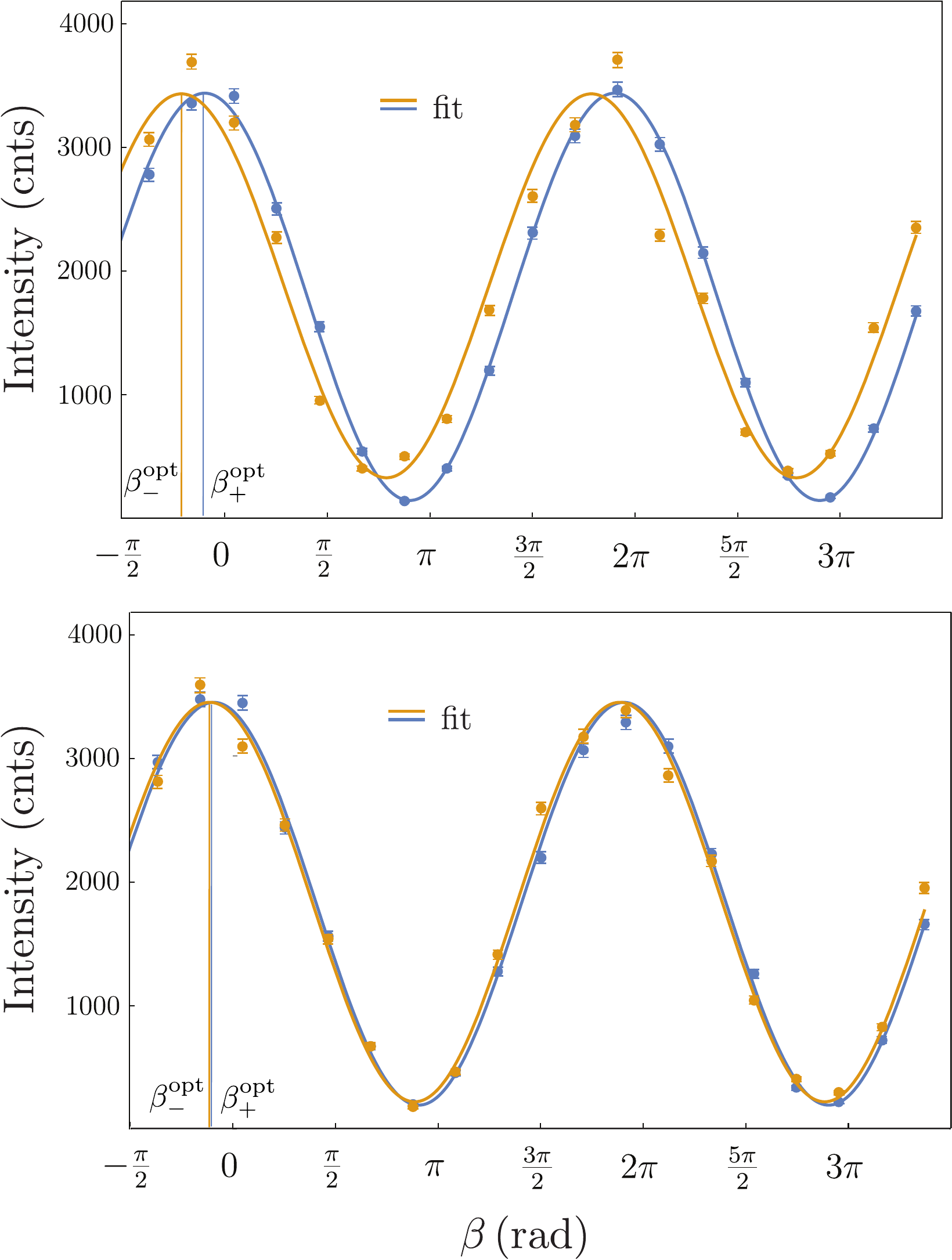}
    \caption{Spin $\ket{\uparrow_x}$ intensities vs. $\beta$ at $\chi=0.04\,\pi$ (top) and at $\chi=1.56\,\pi$ (bottom), measured at the $+$ port (blue) and $-$ port (orange). The optimal compensation angles $\beta^{\rm{opt}}_\pm$ are given by the positions of the maxima (indicated by vertical lines).
    }
    \label{fig:BetaShifts}
\end{figure}

The experiment was carried out at the neutron interferometer instrument S18 at the high-flux reactor of the Institute Laue-Langevin (ILL) in Grenoble, France. A monochromatic beam with mean wavelength $\lambda=\SI{1.91}{\angstrom}$ $(\frac{\delta\lambda}{\lambda}\sim0.02)$ and $\SI{4x8}{mm^2}$ beam cross section was used. The experimental data can be found on the ILL data server \cite{ILL-DATA.3-16-13}.

\subsection{Experimental Results}
\subsubsection{Probabilities}
In order to extract $p_+$ experimentally, the contrast $C$ of the symmetric interferometer with no absorber ($a_1=a_2=\frac{1}{\sqrt{2}}$) has to be measured and used as a reference, as the probability of finding neutrons in the $\ket{+}$ state with $a_1=a_2$ is
\begin{align}
    \vert\braket{+\vert\psi(\chi)}\vert^2 = \frac{1}{2}+C\cos\chi,
\end{align}
Then the measurements are repeated with an asymmetric interferometer realized by an Indium absorber in path 2 ($a_1=\frac{2}{\sqrt{5}}$ and $a_2=\frac{1}{\sqrt{5}}$), and the results are evaluated in relation to the aforementioned contrast. The probability $p_+$ is plotted in Fig.\,\ref{fig:p+} as a function of the phase of initial state $\chi$ (see Appendix A for details).

\begin{figure}[!t]
    \centering
    \includegraphics[width=0.47\textwidth]{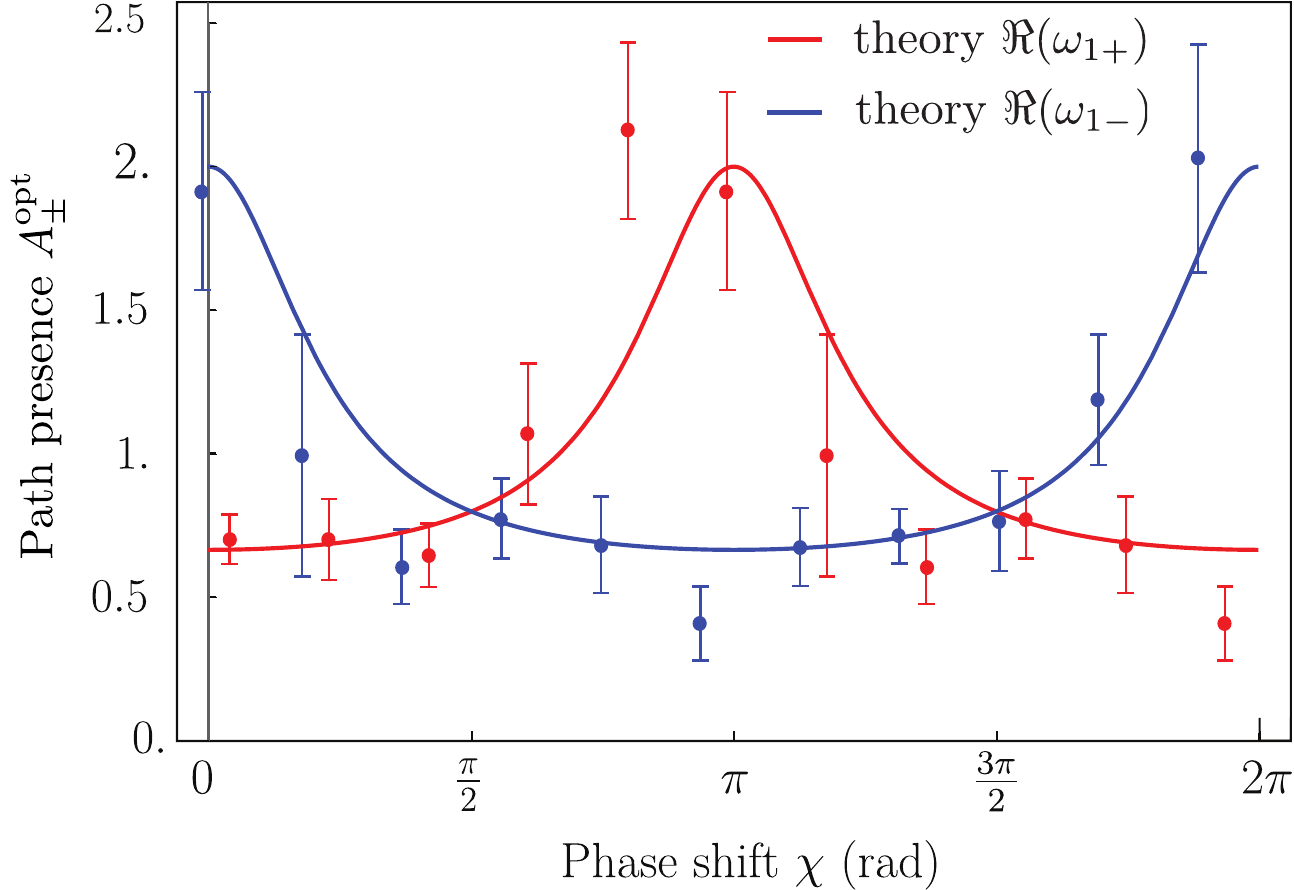}
    \caption{Measured path presence $A^{\text{opt}}_+=\frac{\beta^{\text{opt}}_{+}}{\alpha}$ (red) and $A^{\text{opt}}_-=\frac{\beta^{\text{opt}}_{-}}{\alpha}$ (blue) vs. phase $\chi$ of the initial state, together with theoretical predictions $\Re(\omega_{1\pm})$.}
    \label{fig:ReWV1}
\end{figure}

\begin{figure}[!b]
    \centering
    \includegraphics[width=0.47\textwidth]{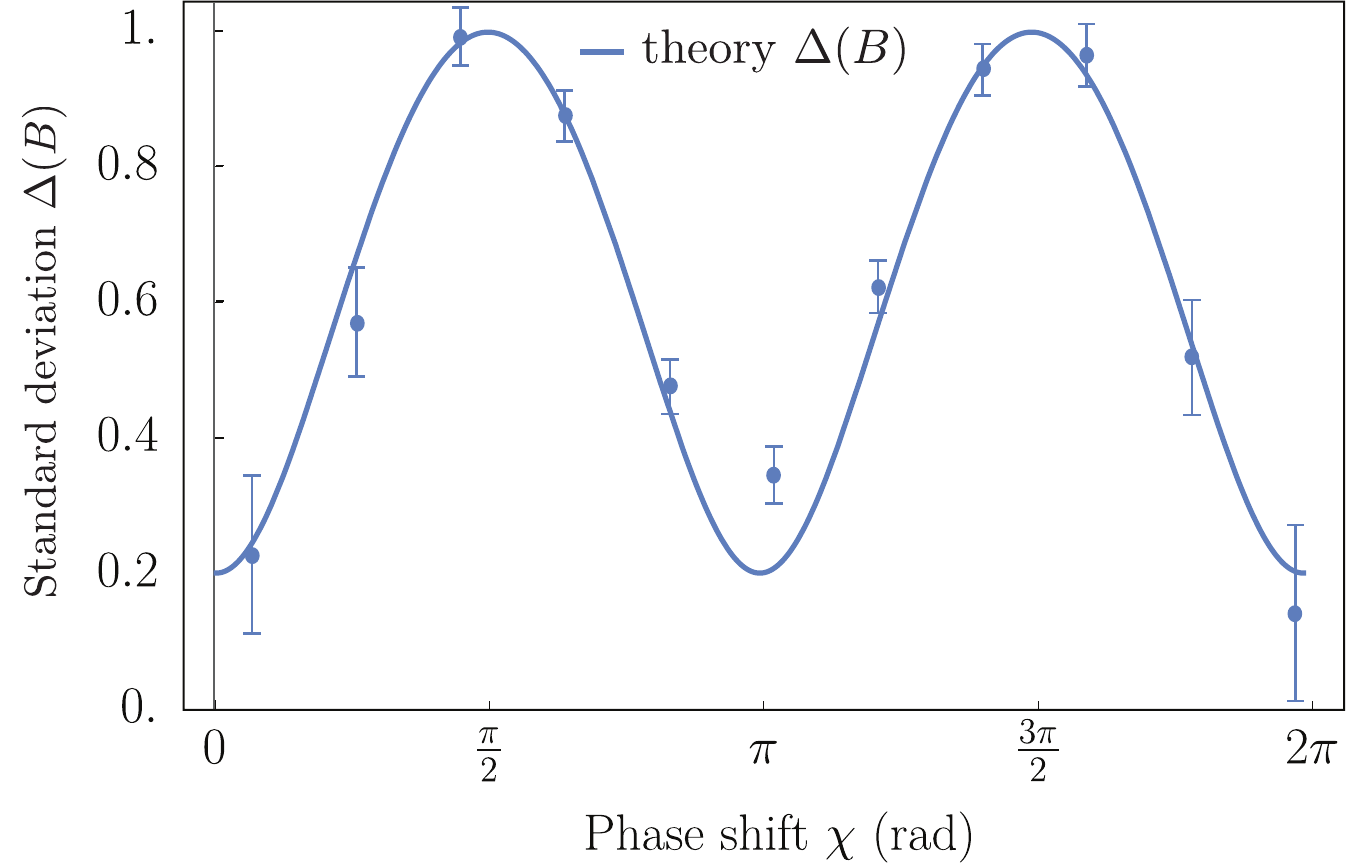}
    \caption{Standard deviation $\Delta(\hat B)$, together with theoretical prediction.}
    \label{fig:DeltaB}
\end{figure}

\subsubsection{The real part of the weak values}
The real part of the weak values $\Re(\omega_{1\pm})$ is given by the optimal estimates $A_\pm^{\text{opt}} = \beta_\pm^{\text{opt}} / \alpha$. The optimal compensation angles $\beta_\pm^{\text{opt}}$ are determined by the positions of the maxima in a $\beta$ scan where the initial polarization is maximally restored, cf. Fig.\,\ref{fig:BetaShifts}. For the plots we have chosen the two phases $\chi$ of the initial state where the optimal compensation angles differ maximally (at $\chi=0.04\,\pi$) and minimally (at $\chi=1.56\,\pi$) respectively in the two output ports. The measured values are for $\chi=0.04\,\pi$: $\beta_+^{\text{meas}}=0.090(7)\,\pi$, $\beta_-^{\text{meas}}=0.191(26)\,\pi$ and for $\chi=1.56\,\pi$: $\beta_+^{\text{meas}}=0.096(10)\,\pi$, $\beta_-^{\text{meas}}=0.096(13)\,\pi$. Fig.\,\ref{fig:ReWV1} shows the obtained optimal estimates $A_\pm^\text{opt}$ as function of $\chi$. They have a $2\pi$ periodicity and range from $\frac{2}{3}$ to 2. For exit $+$ we find a pronounced maximum at $\pi$ where also the errors are largest. Complete compensation is only possible for purely real weak values $w_{1\pm}$, that is for phase shifter setting $\chi=0$ and $\chi=\pi$ \cite{Lemmel2022} (see Appendix C for experimental details of the extraction of path presences). 

\subsubsection{Measurement uncertainty relations}
Experimentally, both $\Delta(\hat B)$ and $\varepsilon(\hat A)$ depend on the values obtained for the output probabilities $p_+$ and $p_-$. Since the probabilities always add up to one, it is possible to use only $p_+$. The uncertainty $\Delta(\hat B)$ can then be expressed directly in terms of the results shown in Fig.\,\ref{fig:p+}
\begin{align}
    \Delta{B} = 2\sqrt{p_+\left(1-p_+\right)},
\end{align}

\begin{figure}[!t]
    \centering
    \includegraphics[width=0.47\textwidth]{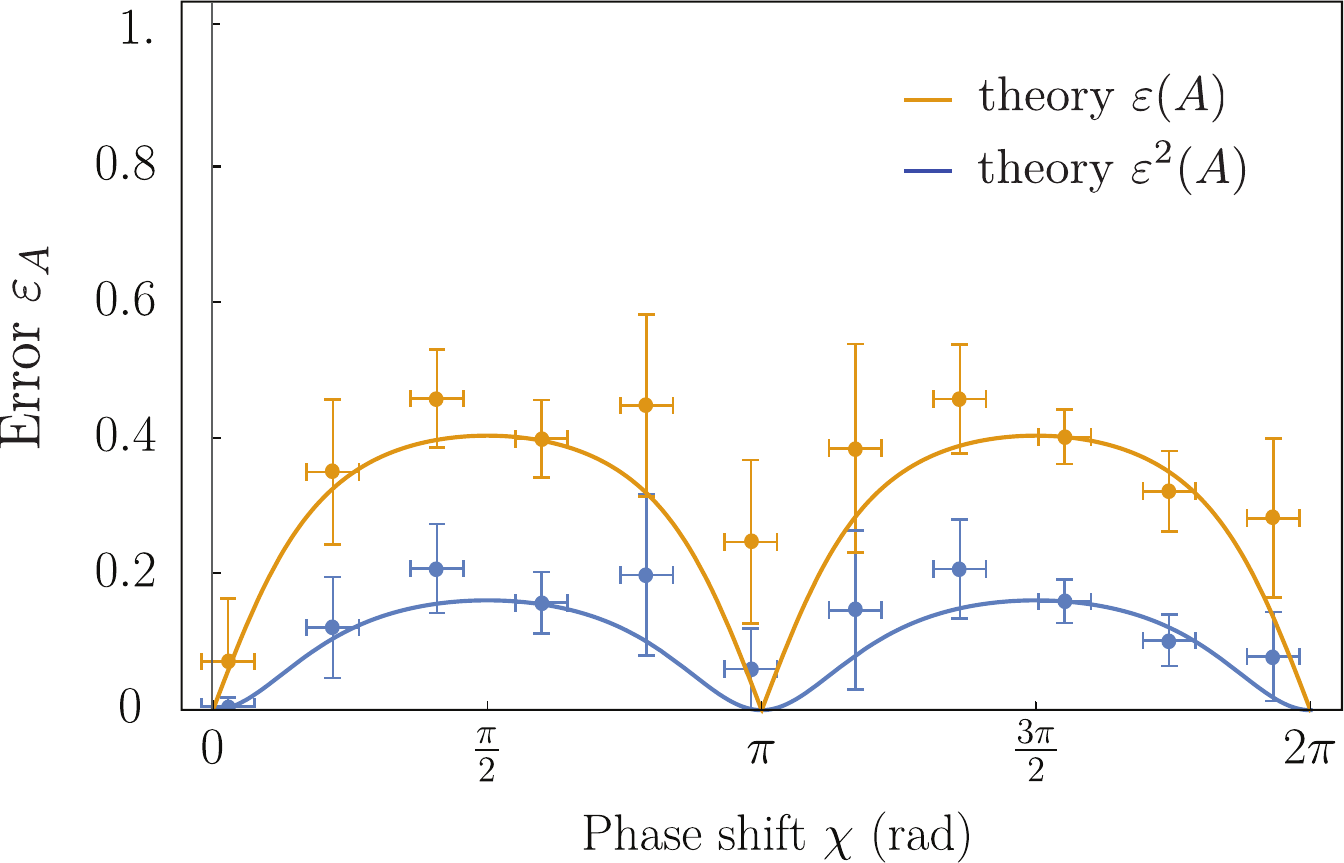}
    \caption{Squared Ozawa Hall Error $\varepsilon^2(\hat A)$ (blue) and Error $\varepsilon(\hat A)$ (orange) vs. phase initial state $\chi$, together with theoretical predictions.}
    \label{fig:epsilon}
\end{figure}

The experimental results are plotted in Fig.\,\ref{fig:DeltaB}. The square of the error $\varepsilon^2(\hat A)$ is obtained by combining the probabilities with the optimal compensation ratios $\frac{\beta^{\text{opt}}_\pm}{\alpha}$. Specifically, the error $\varepsilon^2(\hat A)$ is given by the differences between the compensation ratios and the complex weak values $\omega_{1\pm}$ predicted by the initial statistics of $\hat A$ in the input state $\ket{\psi(\chi)}$, cf. Eq.\,(\ref{eq:error_proj2}),
\begin{align}
    \varepsilon^2(\hat A) = p_+\left\vert w_{1+} - \frac{\beta_+}{\alpha}\right\vert^2 + \left(1-p_+\right)\left\vert w_{1-} - \frac{\beta_-}{\alpha}\right\vert^2.
\end{align}
The experimental results obtained from the data in Fig.\,\ref{fig:p+} and Fig.\,\ref{fig:ReWV1} are plotted in Fig.\,\ref{fig:epsilon}.
We introduce horizontal error bars to account for the fact that the probabilities and the path presences have been measured at slightly different values of $\chi$. The theoretical curve is obtained by replacing the experimental values of $p_\pm$, $\omega_\pm$ and $\beta_\pm/\alpha$ by the theoretical ones given by Eq.\,(\ref{eq:probability}), (\ref{eq:WeakValue}) and (\ref{eq:opt}) respectively.
\begin{align}
  \varepsilon^2(\hat A) = \frac{\sin^2{\chi}}{1-\left(2\,a_1\,a_2\cos{\chi}\right)^2}\left(a_1\,a_2\right)^2
\end{align}

\begin{figure}[!b]
    \centering
    \includegraphics[width=0.47\textwidth]{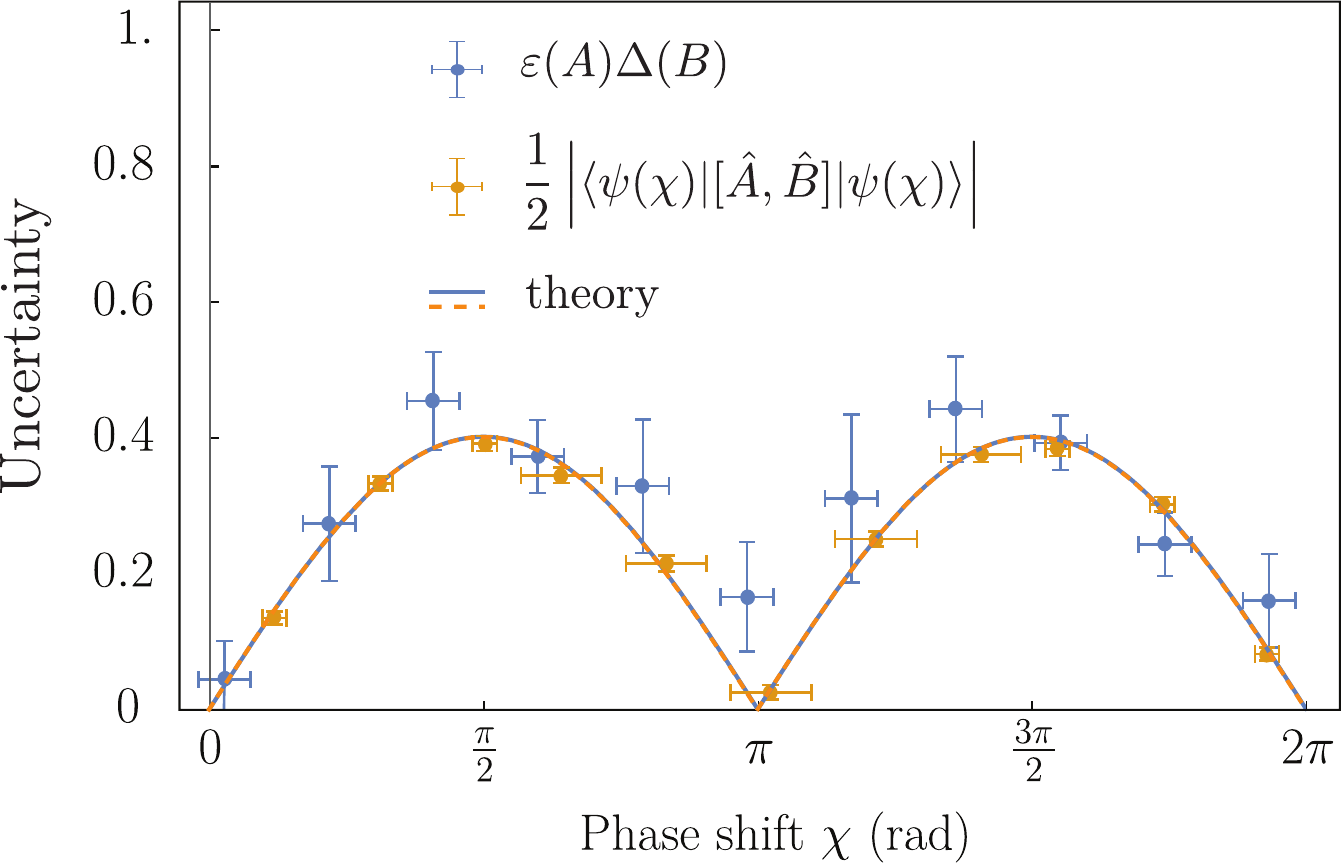}
    \caption{Ozawa's universally valid uncertainty relation in reduce (tight) form. Blue data points represent the LHS and orange data points the RHS of the inequality Eq.(\ref{eq:tightUVUR}). The solid curve shows the theoretical limit where both sides are equal.}
    \label{fig:limit}
\end{figure}

Finally, we are able to compare both sides of the Ozawa relation Eq.\,(\ref{eq:tightUVUR}). The left-hand side is given by $\varepsilon(\hat A)$ and $\Delta(\hat B)$ and the right-hand side by $p_+$ inserted into Eq.\,(\ref{eq:RHS}). The result is plotted in Fig.\,\ref{fig:limit}. 
As expected, we obtain a \emph{tight} relation and the error vanishes completely for $\chi$ equal to a multiple of $\pi$ where the weak values are real. Since the measured path presences differ from theoretical prediction at some phase shifts $\chi$ (see Fig.\,\ref{fig:ReWV1}) and the value of $\varepsilon(\hat A)$ is increased by additional noise, the measured points for $\varepsilon(\hat A)\,\Delta(\hat B)$ are slightly above the theoretical prediction.

\section{Conclusion}
In this work we compare measurements of the Ozawa-Hall error and the measured lower bound of Ozawa's universally valid uncertainty relations as function of initial states's phase $\chi$ for observables $\hat{A}=\hat{\Pi}_1$ and $\hat{B}=\hat{\sigma}_x$. It is demonstrated that the uncertainty relation is tightly fulfilled for all states $\ket{\psi(\chi)}$. The measurement error $\varepsilon(A)$ vanishes completely for phase settings where the weak value is purely real (imaginary part is zero). Since the lower bound of the uncertainty relation is given by the gradient of the interference fringe the bound also vanishes at the same phases, namely $\chi=0$ and $\chi=\pi$. This illustrates that in Ozawa’s theory of measurement errors the imaginary part of the weak value is associated with the error. Measurements also of the imaginary part of the weak value by applying modified feedback compensation is topic of forthcoming experiments. A possible application of the imaginary part of the weak value would be a direct determination of the uncertainty bound as presented in \cite{Wagner21}, where a connection between the commutator relation and imaginary part of the weak value is established.

Furthermore, we want to point out that in the special case studied here, with zero disturbance $\eta(\hat B)$, the in general sub-optimal Ozawa uncertainty relation (Eq.\,(\ref{eq:Ozawa})) coincides with the tighter Branciard relation \cite{Branciard13}.

\begin{acknowledgments} 
This work was supported by the Austrian science fund (FWF) Projects No. P 34239, P 30677, and P 34105. HFH was supported by ERATO, Japan Science and Technology Agency (JPMJER2402). We acknowledge the ongoing hospitality and ongoing support of the ILL, Grenoble (France). The authors acknowledge TU Wien Bibliothek for financial support through its Open Access Funding Program.
\end{acknowledgments}
\newpage

\quad \quad\quad \quad\quad \quad\quad {\textbf{APPENDICES}}
\appendix
\section{Probabilities}
A conventional \emph{Which-Way} measurement, realized by placing a detector in path 1 or 2 directly behind the first plate (see Fig.\,\ref{fig:setup}), results in $p_1=0.799\,4\pm0.002\,9$ and $p_2=0.200\,6\pm0.002\,9$. The imbalance is due to the partial absorber  placed in path 2. This corresponds to an amplitude ratio $a_1:a_2=0.894\,1\pm0.001\,6:0.447\,9\pm0.003\,3$ which is a very good approximation of $\frac{2}{\sqrt{5}}:\frac{1}{\sqrt{5}}$ or  $a_1:a_2=2:1$.

The range of the probability $p_+$ can be evaluated as the ratio between the contrasts obtained from the interferograms with ($C_{2:1}$) and without ($C_{1:1}$) an absorber in path 2, namely $\text{max}\,(p_+)-\text{min}\,(p_-)=C_{2:1}/C_{1:1}$. To determine the contrasts $C_{2:1}$ and $C_{1:1}$ we have selected the phase shifter range with highest contrast and divided the interferograms by the sum of the O+H counts for normalization.

\section{Lower bound of uncertainty}
It is possible to determine $\frac{1}{2}\left\vert\braket{[\hat{A},\hat{B}]}\right\vert$ (lower bound from uncertainty relation from Eq.\,(\ref{eq:tightUVUR})) from $p_+$ by shifting the phase as $\chi\rightarrow\chi\mp\frac{\pi}{2}$. Namely via
\begin{align}
    \frac{1}{2}\left\vert\bra{\psi}[\hat{A},\hat{B}]\ket{\psi}\right\vert = \frac{1}{2}\left\vert\bra{\psi}[\hat{\Pi}_1,\hat{\sigma}_x]\ket{\psi}\right\vert =\nonumber\\
    \frac{1}{2}\left\vert\bra{\psi}\I\,\hat{\sigma}_y\ket{\psi}\right\vert = \frac{1}{2}\left\vert\bra{\psi}\begin{pmatrix}1&0\\0&\I\end{pmatrix}\I\,\hat{\sigma}_x\begin{pmatrix}1&0\\0&-\I\end{pmatrix}\ket{\psi}\right\vert =\nonumber\\
    \frac{1}{2}\left\vert\bra{\psi\left(\chi-\frac{\pi}{2}\right)}\I\,\hat{\sigma}_x\ket{\psi\left(\chi-\frac{\pi}{2}\right)}\right\vert =\nonumber\\
    \frac{1}{2}\left\vert\,p_+\left(\chi-\frac{\pi}{2}\right)-p_-\left(\chi-\frac{\pi}{2}\right)\right\vert =\nonumber\\
    \frac{1}{2}\left\vert\,p_+\left(\chi-\frac{\pi}{2}\right)-p_+\left(\chi+\frac{\pi}{2}\right)\right\vert.
    \label{eq:appRHS}
\end{align}

\section{Intensities of an ideal interferometer}
Let us consider a Mach-Zehnder interferometer setup like the one shown in Fig.\,\ref{fig:setup} (b).  The first beam splitter is asymmetric with amplitudes $a_1$ and $a_2$ for path 1 and 2, respectively. Together with a subsequent relative phase shift $\chi$ between path 1 and 2 the initial path state is expressed as
    \begin{align}
    \ket{\psi(\chi)} = a_1\ket{1}+a_2\e^{\I\chi}\ket{2}.
\end{align}
Since we prepare our initially spin in in $+x$-direction with spin state $\ket{\uparrow_x}=1/\sqrt{2}(\ket{\uparrow_z}+\ket{\downarrow_z})$ the total initial state (consisting of path and spin state) reads
    \begin{align}
    \ket{\Psi_{\rm in}^\chi} =\ket{\psi(\chi)}  \ket{\uparrow_x}.
\end{align}
Next we rotate the spin by a small angle $\alpha$ about the $z$ axis only in path 1. The rotation is expressed by the operator $\hat U_z^{\alpha}$ acting only in path 1 or, equivalently, by the operator $\hat U_{z1}^{\alpha}$ acting on the total state of both paths and spin.
\begin{subequations}
\begin{alignat}{1}
  \hat U_{z1}^\alpha &= \exp \left(-\frac \I 2 \alpha \, \hat \sigma_z \,  \hat \Pi_1 \right) = 
 \hat \Pi_1  \hat U_{z}^\alpha + \hat \Pi_2  \\
  \hat U_{z}^\alpha &=  \exp \left(-\frac \I 2 \alpha \, \hat \sigma_z\right) = {1\!\!1} \cos \frac \alpha 2 - \I \hat \sigma_z \sin \frac \alpha 2
\end{alignat}
\end{subequations}
where $ \hat \Pi_1$ and $ \hat \Pi_2$ denote the path projection operators of path 1 and 2 respectively. The total state after spin rotation $\alpha$ reads
\begin{alignat}{1}
  |\Psi^\alpha\rangle &=  \hat U_{z1}^\alpha |\Psi^\chi_{\text{in}}\rangle = \hat U_{z1}^\alpha   |\psi(\chi)\rangle     |\uparrow_x\rangle.
\end{alignat}
The states of the two exit beams of the interferometer are given by the projection onto the exit states $|\pm\rangle$ respectively
\begin{alignat}{1}
  |\Psi^{\rm out}_\pm\rangle &=|\pm\rangle  \langle\pm |\Psi^\alpha\rangle =  \langle\pm | \hat U_{z1}^\alpha  |\psi(\chi)\rangle \; |\pm\rangle    |\uparrow_x\rangle,
\end{alignat}
with $\ket{\pm} = \tfrac{1}{\sqrt{2}}(\ket{1}\pm\ket{2})$. In the exit beams we compensate the rotation of the spin by rotating it back (again about the $z$-axes) by an angle $\beta_+$ in the $|+\rangle$ port and $\beta_-$ in the $|-\rangle$ port applying a compensation operation $\hat U_z^{\beta\pm}$. The final state at the O-beam ($|+\rangle$ port) reads
\begin{align}
\begin{aligned}
        \ket{\Psi_+^\beta} &= {\langle\pm | \hat U_z^{\beta\pm} \hat U_{z1}^\alpha  |\psi(\chi)\rangle \; |\pm\rangle    |\uparrow_x\rangle} \\ &= \frac{
    a_1\e^{-\I\frac{\alpha+\beta}2 \hat{\sigma}_z} \ket{\uparrow_x}+
    a_2\e^{\I\chi}\e^{-\I\frac{\beta}2 \hat{\sigma}_z} \ket{\uparrow_x}}{\sqrt{2}}\ket{+} \\
        &= \tfrac{1}{\sqrt{2}}\left[\left(a_1\cos{\left(\tfrac{\alpha+\beta}{2}\right)}+a_2\e^{\I\chi} \cos{\tfrac{\beta}{2}}\right)\ket{\uparrow_x}  \right. \\&-\I \left.\left(a_1\sin{\left(\tfrac{\alpha+\beta}{2}\right)}+a_2\e^{\I\chi} \sin{\tfrac{\beta}{2}}\right)\ket{\downarrow_x} \right]\ket{+}.
    \end{aligned}
    \label{eq:psi+}
\end{align}
In the case of a perfect interferometer, the expected intensities after projecting on the states $\ket{\uparrow_x}$ or $\ket{\downarrow_x}$ can be simply evaluated as
\begin{align}
    I_{+x}&= |\braket{\uparrow_x|\Psi_+^\beta}|^2 \\
    I_{-x}&= |\braket{\downarrow_x|\Psi_+^\beta}|^2,
\end{align}
which are proportional to the probabilities $p_+^{\uparrow x}$ and $p_+^{\downarrow x}$ from the Mach-Zehnder interferometer scheme depicted in Fig.\,\ref{fig:setup} (b).

\section{Extraction of path presences $\omega_{1\pm}$ and optimal estimate $A^{\rm{opt}}_\pm $}
Using the states of the two outgoing beams from the $\ket{+}$ and $\ket{-}$ port the weak values of the path projection operator onto path 1 are given by 
\begin{align}
    \omega_{1\pm}(\chi) \overset{\text{Eq.}\,(\ref{eq:WeakValue})}{=} \frac{1}{1\pm\frac{a_2}{a_1}\e^{\I\chi}} = \nonumber\\
    = a_1^2\mp\frac{a_1\,a_2\left(\cos{\chi}\left(a_1^2-a_2^2\right)+\I\sin\chi\right)}{1\pm2\,a_1\,a_2\cos{\chi}}
\end{align}
respectively. Experimentally the path presences $\omega_{1\pm}$ are obtained using \emph{feedback compensation} with a small (fixed) interaction strength of $\alpha=\pi/8$. In order to reproduce the theoretical predictions accurately the interfering / non interfering parts of the neutron's wave function in the interferometer have to be taken into account in detail, which is explained in the subsequent sections.

As shown in detail in \cite{Lemmel2022}, in the limit of small interaction strength $\alpha$ it is possible to implement the expansion
\begin{align}
    \hat U_z^{\beta\pm} \hat U_{z1}^\alpha &= \exp\left[ -\frac{i\alpha}{2}\hat \sigma_z \left(\hat \Pi_1-\frac{\beta_\pm}{\alpha}\right)\right] \\& \approx 1-\frac{i\alpha}{2}\hat \sigma_z \left(\hat \Pi_1-\frac{\beta_\pm}{\alpha}\right)
\end{align}
and, consequently, Eq.\,(\ref{eq:psi+}) becomes
\begin{align}
\begin{aligned}
        \ket{\Psi_+^\beta} &= {\langle\pm | \hat U_z^{\beta\pm} \hat U_{z1}^\alpha  |\psi(\chi)\rangle \; |\pm\rangle    |\uparrow_x\rangle}\\
        &\approx {\langle\pm | \left[1-\frac{i\alpha}{2}\hat \sigma_z \left(\hat \Pi_1-\frac{\beta_\pm}{\alpha}\right)\right]  |\psi(\chi)\rangle \; |\pm\rangle    |\uparrow_x\rangle}
        \\&\approx \langle\pm | \psi(\chi)\rangle  \left[1-\frac{i\alpha}{2}\hat \sigma_z \left(\omega_{1\pm}(\chi)-\frac{\beta_\pm}{\alpha}\right)\right] \; |\pm\rangle    |\uparrow_x\rangle \, .
    \end{aligned}
    \label{eq:psi+_weak}
\end{align}
From Eq.\,\ref{eq:psi+_weak} we can see that optimal compensation happens for 
\begin{align}
    \beta_\pm = \beta^{\rm{opt}}_\pm = \alpha \,   \Re\big(\omega_{1\pm} (\chi)\big) \, .
\end{align}
Hence, in the limit of weak coupling (small $\alpha$) the weak value of the path projection operator is given by the ratio of the spin rotation angle $\alpha$ and the optimal compensation $\beta_{0\pm}$
\begin{alignat}{1}
 \Re\big(\omega_{1\pm} (\chi)\big)&\approxeq \beta^{\rm{opt}}_{\pm} /  \alpha:=A^{\rm{opt}}_\pm ,
\end{alignat}
where $A^{\rm{opt}}_\pm $ is referred to as \emph{optimal estimate} of the path presence $\omega_{1\pm} (\chi)$.

\section{Intensities of a non-ideal interferometer}
In the case of a non-ideal interferometer, the neutrons might not undergo path interference, and this would result in a percentage of non interfering (n.i.) neutrons that will reach the detector. Therefore, the description of the measured intensity has to take into account both the non interfering  and the interfering components of the intensity. The former can be described by considering the contributions from path 1 and path 2 separately. The neutrons coming from path 1 will go through all the spin rotations $\alpha$ and $\beta$, while the ones coming from path 2 will only experience the spin rotations $\beta$. The phase shifter will have no effect since the neutrons are not interfering. By rearranging Eq.\,(\ref{eq:psi+}) we can isolate the terms relative to the contribution from each single path:
\begin{align}
    \begin{aligned}
        &\ket{\Psi_+^\beta} = 
        \tfrac{1}{\sqrt{2}}\left[a_1\left(\cos{\left(\tfrac{\alpha+\beta}{2}\right)}\right) + a_2\e^{\I\chi}\left(\cos{\tfrac{\beta}{2}}\right)\right]\ket{\uparrow_x}\ket{+} +\\&+ \tfrac{1}{\sqrt{2}}\left[a_1\left(-\I\sin{\left(\tfrac{\alpha+\beta}{2}\right)}\right) + a_2\e^{\I\chi}\left(-\I\sin{\tfrac{\beta}{2}}\right)\right]\ket{\downarrow_x}\ket{+}.
    \end{aligned}
\end{align}
Consequently, the non interfering intensity coming from path 1 will be
\begin{align}
     I^{\text{n.i.}}_{1,+x} = \frac{a_1^2}{2}\left|\cos{\left(\tfrac{\alpha+\beta}{2}\right)}\right|^2, \nonumber\\
     I^{\text{n.i.}}_{1,-x} = \frac{a_1^2}{2}\left|-\I\sin{\left(\tfrac{\alpha+\beta}{2}\right)}\right|^2,
\end{align}
and the contribution from path 2 will be
\begin{align}
    I^{\text{n.i.}}_{2,+x} = \frac{a_2^2}{2}\left|\cos{\tfrac{\beta}{2}}\right|^2, \nonumber\\
    I^{\text{n.i.}}_{2,-x} = \frac{a_2^2}{2}\left|-\I \sin{\tfrac{\beta}{2}}\right|^2.
\end{align}
The total contribution from the non interfering neutrons can then be expressed as:
\begin{align}
    I^{\text{n.i.}}_{+x}= I^{\text{n.i.}}_{1,+x} + I^{\text{n.i.}}_{2,+x}, \nonumber\\
    I^{\text{n.i.}}_{-x}= I^{\text{n.i.}}_{1,-x} + I^{\text{n.i.}}_{2,-x}.
\end{align}
The measured intensity relative to the $+x$ component is then equal to
\begin{align}
    I^\text{{meas.}}_{+x}= C\, I^{\text{in.}}_{+x} + (1-C) \, I^{\text{n.i.}}_{+x}
    \label{eq:correction}
\end{align}
where $C$ is the measured contrast from an interferogram. We obtained $I^{\text{in.}}_{+x}$ according to Eq.\,(\ref{eq:correction}). This gives the observed results for the $\beta$ shifts ($\Re(\omega_{1\pm}))$ from the main text.


%

\end{document}